\documentclass[%
twocolumn,superscriptaddress,
 amsmath,amssymb,
 aps,
prc,
]{revtex4-1}
\usepackage{mathrsfs,amsmath,amssymb,amsthm}
\usepackage{amssymb,fmtcount}
\usepackage{graphicx,epsfig,latexsym,overpic,amssymb,color}
\usepackage{threeparttable}
\usepackage{amsmath}
\usepackage{float}
\usepackage{changes}

\everymath{\displaystyle}
\usepackage{graphicx}% Include figure files
\usepackage{dcolumn}
\begin{document}

%\preprint{}

\title{ Core screening effect in knockout reactions
}
%  Obstruction

\author{Shichang Li}
%\thanks{}
\affiliation{
State Key Laboratory of Nuclear Physics and Technology, School of Physics,
Peking University, Beijing 100871, China
}
\author{J.C. Pei}\email{peij@pku.edu.cn}
\affiliation{
State Key Laboratory of Nuclear Physics and Technology, School of Physics,
Peking University, Beijing 100871, China
}
\affiliation{
Southern Center for Nuclear-Science Theory (SCNT), Institute of Modern Physics, Chinese Academy of Sciences, Huizhou 516000,  China
}
\author{D.Y. Pang}
\affiliation{
School of Physics, Beihang University, Beijing 100191, China
}

\date{\today}  %

\begin{abstract}
The systematic quenching of spectroscopic factors in terms of separation energy asymmetry in single-nucleon knockout reactions remains a puzzle.
We propose a core screening effect to consider the hindrance when strongly bound nucleons in the projectile nucleus are removed by
the heavy-ion target.
The core screening effect is simulated as a density dependent suppression of single-particle wave functions inside the core of projectile.
Our study shows that the parameterized core screening effect can significantly reduce the isospin  dependence of quenching factors,
offering insights into nuclear reaction mechanisms.
\end{abstract}

%\keywords{}

\maketitle

% References should be done using the \cite, \Ref.~and \label commands
% \section{The NN potential and local projection}

\section{Introduction}

In the past decade, systematic studies on single-nucleon knockout reactions induced by heavy ions have demonstrated  significant  discrepancies between experimental and theoretical results~\cite{PhysRevC.77.044306,PhysRevC.90.057602,PhysRevC.103.054610} .
Such discrepancies are expressed as quenching factor $R_s=\sigma_{exp}/\sigma_{th}$, which increase
in terms of the separation energy asymmetry $\Delta S$. Specifically, $\Delta S$=$S_n$-$S_p$ for neutron removal and
$\Delta S$=$S_p$-$S_n$ for proton removal.
It is still puzzling that $R_s$ approaches 1 for knockout of weakly-bound nucleons but decreases to 0.2$\thicksim$0.4 for knockout of deeply-bound nucleons~\cite{PhysRevC.103.054610}.
 However, such dependence of quenching factors on $\Delta S$ is not observed in  transfer reactions\cite{PhysRevLett.131.212503,PhysRevLett.110.122503,PhysRevC.73.044608}, $(p,pN)$ reactions\cite{GOMEZRAMOS2018511,PhysRevC.100.064604}, or electron-induced knockout reactions\cite{KRAMER2001267}.
This indicates that the modeling of knockout reactions involving heavy-ion projectiles is more complex than expected.

There has been extensive measurements of  spectroscopic factors of unstable nuclei using light nuclei such as $^9$Be and $^{12}$C as targets via one-nucleon removal reactions \cite{AUMANN2021103847,PhysRevC.91.041302,PhysRevC.65.061601,PhysRevC.90.037601,PhysRevC.98.024306,PhysRevC.79.024616},
 which is useful for understandings of the evolution of shell structures.
These reactions are conventionally described using the sudden approximation and the eikonal model.
The higher-energy measurements indicate that the approximately linear reduction in $R_s$ is not related to the breakdown of the sudden approximation~\cite{PhysRevC.103.054610}.
The eikonal model is a good semi-classical approximation to study halo nuclei at high and intermediate energies \cite{J.A.Tostevin_1999,annurev:/content/journals/10.1146/annurev.nucl.53.041002.110406,HUSSEIN1985124}.
However, the observation of significant dissipative core-target interaction in the knockout reaction with a composite target
implies new reaction mechanisms beyond the eikonal model are needed~\cite{sun}.

The quenching phenomena in isospin asymmetric nuclei has attracted strong interests.
Recently the core destruction effect is proposed to explain the significant reduction of cross sections for removal of deeply bound nucleons\cite{GOMEZRAMOS2023138284}.
However, another study did not find further reduction of cross section due to the core destruction effect with increasing binding energy of the removed nucleon\cite{BERTULANI2023138250}.
In addition, {\it ab initio} wave functions based on no-core shell model~\cite{PhysRevC.105.024613} and Gamow shell model with continuum coupling~\cite{jianguo-plb,gamow} have been adopted in the reaction model, but the quenching phenomena remains an issue.
The short-range correlation effects dominated by $np$ pairs within the dynamical Li\`{e}ge intranuclear-cascade (INCL) model
is promising to explain the $\Delta S$ dependence~\cite{RODRIGUEZSANCHEZ2024138559}, but explanations on proton-rich nuclei are needed.

 In this work, we propose a parameterized core screening effect to consider the hindrance when the heavy ion target interacts with  deeply bound nucleons.
 The hindrance is understandable as the heavy ions penetrate into the projectile core.
  This screen effect is realized by including a density dependent in-medium wave functions of valence nucleons within the reaction model.
  Then deeply bound nucleons inside the core would feel more screening effect than weakly bound nucleons.
 Actually the  screening effect is a result of the complex short-range interactions  between two composite systems such as the Pauli blocking effect.
 For example, the Pauli blocking effect is crucial in descriptions of $\alpha$-particle embedded in nuclear matter and the fusion hindrance, which can be taken into account by a density dependent formalism~\cite{xu1,xu2}.
 It is also known that the in-medium nucleon-nucleon interaction in transport model would be reduced in heavy ion collisions~\cite{qfli}.
 Our numerical results indicate that, with the inclusion of the core screen effect, the dependence of the quenching factor on isospin asymmetry is significantly reduced.

%%%%
\section{Method}

The knockout reaction is a rapid process with minimal rearrangement or compound formation before the nucleon is knocked out.
The theoretical description of the single-nucleon knockout reaction cross section can be decomposed into two main components: the nuclear structure factor $C^2S$ and the
single-particle cross section $\sigma_{sp}$ \cite{PhysRevC.77.044306,AUMANN2021103847,TOSTEVIN2001320}.
\begin{equation}
    \sigma=\sum_{J^{\pi},nlj}(\frac{A}{A-1})^N C^2S(J^{\pi},nlj)\sigma_{sp}(nlj)
\end{equation}
The spectroscopic factor $C^2S$ is a crucial input for describing the occupancy of a specific single-particle state with quantum numbers $nlj$.
The spectroscopic factor is conventionally calculated by shell model with many-body correlations in a harmonica oscillator basis. The single-particle
 cross-section $\sigma_{sp}$ is the sum of the contributions from the stripping cross-section $\sigma_{sp}^{str}$ and the diffraction dissociation cross-section $\sigma_{sp}^{dif}$\cite{HUSSEIN1985124,annurev:/content/journals/10.1146/annurev.nucl.53.041002.110406,PhysRevC.46.2638}.
The stripping cross section is dominated and will be studied in detail in this work.
\begin{equation}
    \sigma_{sp}^{str}=\frac{1}{2j+1}\iint d\vec{b_n}d\vec{r}(1-|S_n(b_n)|^2)|S_c(b_c)|^2|\psi_{nljm}(\vec{r})|^2
    \label{eq:str}
\end{equation}
$S_c$ and $S_n$ represent the $S$-matrices for the scattering of the core (or nucleon) with the target nucleus at collision parameters $b_c$ and $b_n$.
$\psi_{nljm}$ denotes the wave function of the removed nucleon in the target nucleus with quantum numbers $nljm$.
This formula is derived from the  perspective of survival probability.
On the one hand, the core must be detected intact after stripping, with a survival probability of $|S_c|^2$.
On the other hand, the nucleon is removed with a probability of $1-|S_n|^2$.
Combining these two probabilities results in the inelastic break-up, corresponding to the stripping cross-section \cite{HUSSEIN1985124}.
The core destruction effect is proposed to modify the core survival probability $|S_c|^2$, but can not resolve the
systematic quenching puzzle~\cite{BERTULANI2023138250}.

\begin{figure}[!htbp]
	\centering
	\includegraphics[width=0.4\textwidth]{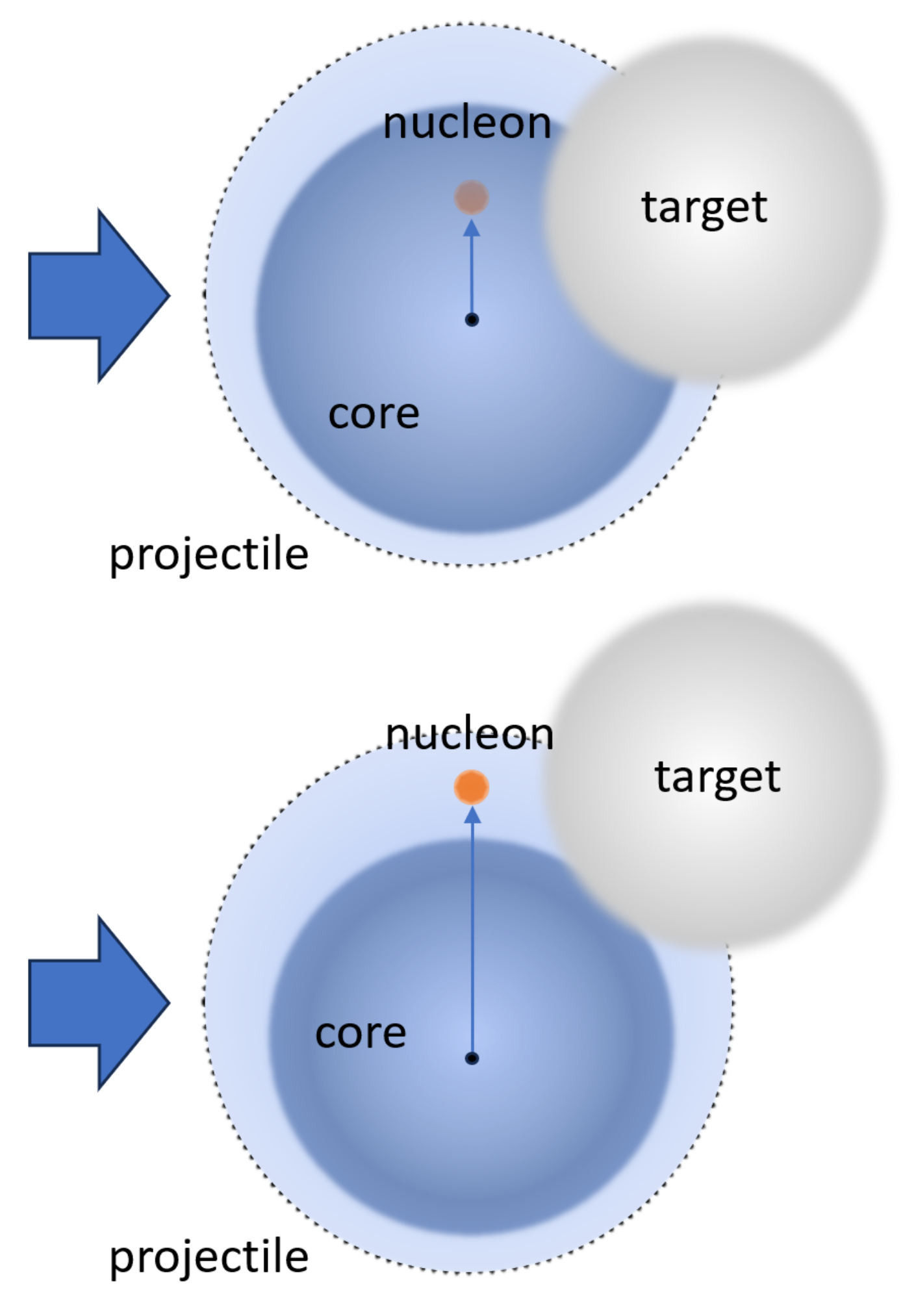}	
	\caption{
    Illustration of the core screening effect of valence nucleon in knockout reaction by using heavy ion target.
    The upper panel shows the knockout of a deeply bound nucleon, in which the core screening effect is significant.
    The lower panel shows the knockout of a weakly bound nucleon, in which the core screening effect is not important. }
	\label{fig_screen}%
\end{figure}
%核心屏蔽效应示意图,上图表示被敲出核子束缚比较深的情况,下图表示被敲除盒子束缚比较弱的情况

In modelings of knockout reactions, the three-body interaction between the core, the valence nucleon and the target nucleus are usually omitted \cite{HUSSEIN1985124,J.A.Tostevin_1999}, which could impact the reaction dynamics significantly. For the interaction between the valence nucleon and the target, the core screening effect could play a role based on the Pauli blocking effect.
The core screening effect is sensitive to the spatial distributions of the wavefunctions.
Thus the reaction involving deeply bound wave functions would be much suppressed by the in-medium core-screening effect.
While the weakly bound nucleons at nuclear surface should feel no screening effect.
In addition, such a core screening is dependent on the target nucleus and should be small for the proton target.
This picture is understandable but has not been examined before.

It is difficult to calculate the Pauli blocking and in-medium effects between two composite systems in reaction models.
This is a general problem such as the descriptions of the short-range nuclear force considering the quark motion  within nucleons,
and the short-range interaction between two atoms with electronic orbitals.
It is reasonable to consider the spatial density distribution as the most crucial degree of freedom in the Pauli blocking and in-medium effects.
Here we propose a parameterized density dependence $f(r)$ to simulate the core screening effect, which applies to the wave functions of valence nucleons in Eq.(\ref{eq:str}).
\begin{equation}
\begin{array}{l}
    f_{}(r)=1-\alpha(\frac{\rho(r)}{\rho_0})^{\gamma} \vspace{5pt}\\
    |\psi_{nljm}(r)|^2 \rightarrow |f(r)\psi_{nljm}(r)|^2 \\
    \label{eq:fs}
\end{array}
\end{equation}
This effective form of density dependence involves two parameters,
which is taken from the form of density dependent pairing interaction~\cite{pair1,pair2}.
Similarly, to simulate the Pauli blocking effect in $\alpha$-cluster embedded in nuclear matter,
a density dependent form has been adopted with three parameters~\cite{xu1,xu2}.
The core destruction in Ref.~\cite{GOMEZRAMOS2023138284} also invokes an effective non-local density.
In Eq.(\ref{eq:fs}), the parameters $\alpha$ and $\gamma$ are introduced to describe the characteristics of the screen effect, while $\rho(r)$ represents the total density of the remaining core. Here, $\rho_0$ denotes the maximum value of $\rho(r)$.

\section{Results}

\begin{figure}
	\centering
	\includegraphics[width=0.4\textwidth]{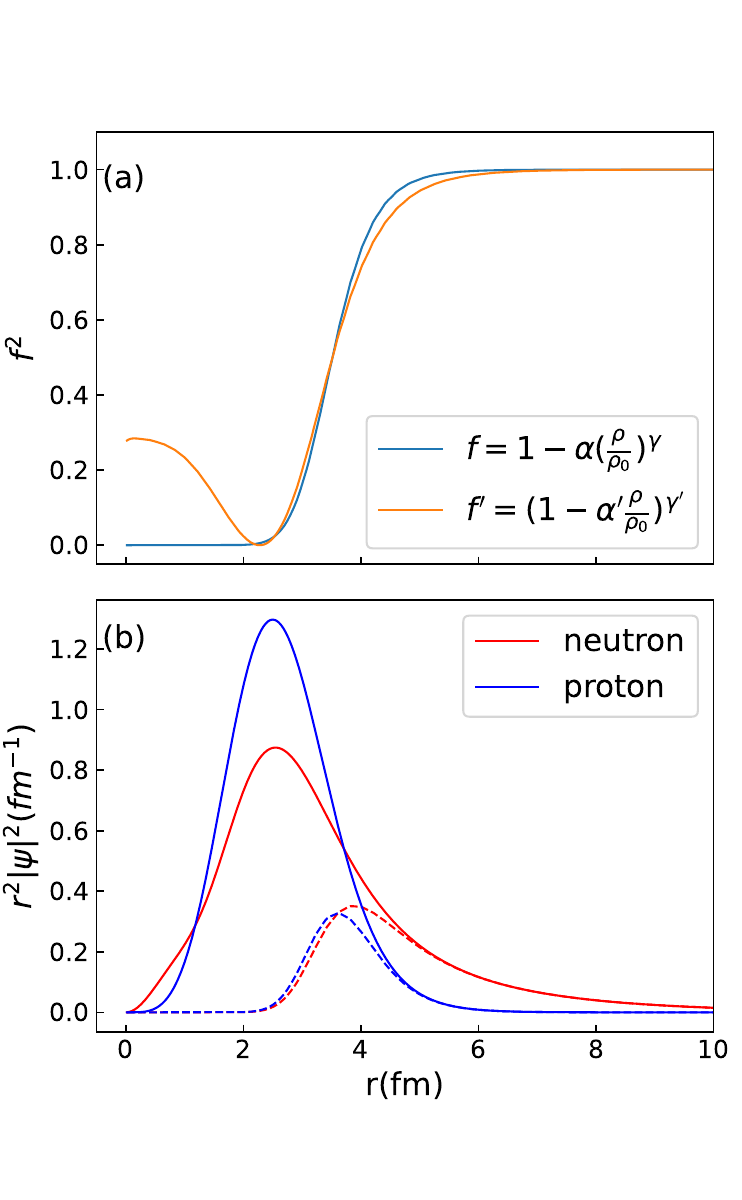}	
	\caption{
Two different density dependent core screening factors are shown in (a).
The wave functions of the removal proton or neutron in $^{15}$C are shown in (b), with (dashed line) or without (solid line) the core screening.
}
	\label{sf}%
\end{figure}
%屏蔽因子随着半径的变化

\begin{figure}
	\centering
	\includegraphics[width=0.4\textwidth]{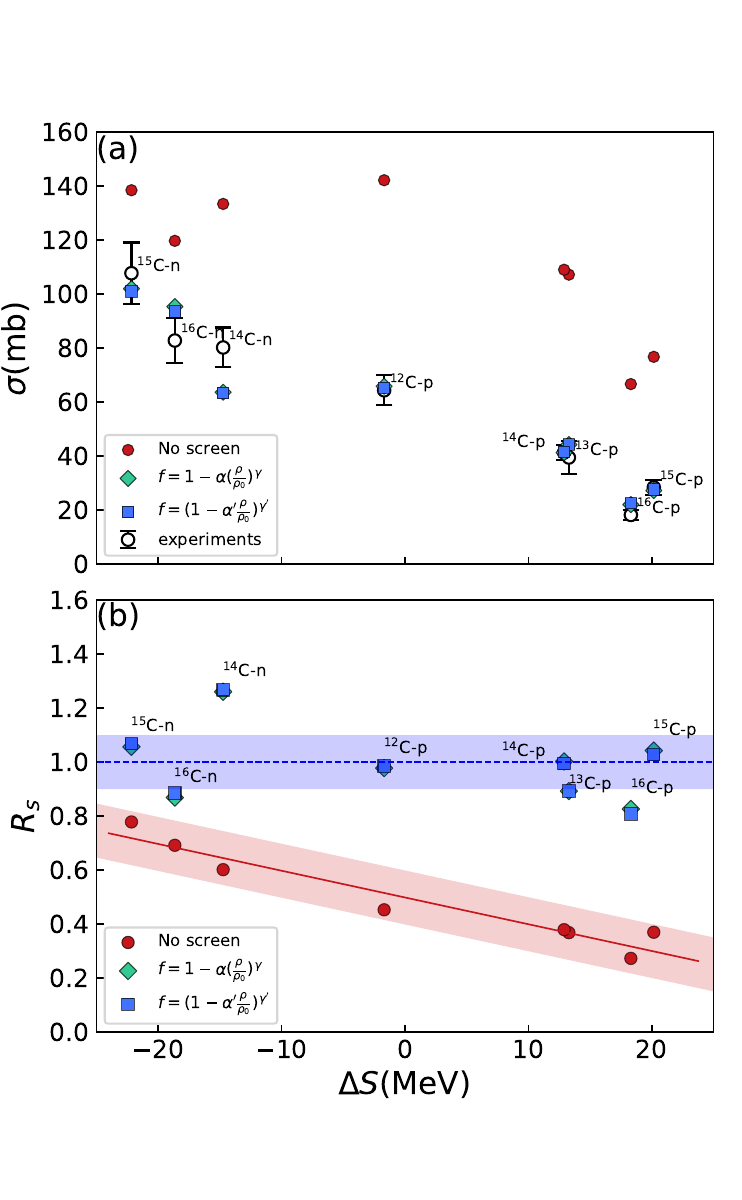}	
	\caption{
   The calculated one-nucleon knockout cross sections of carbon isotopes as a function of $\Delta S$, as shown in (a).
   The cross sections with two different core screening factors are compared with the experimental data~\cite{PhysRevC.110.014603,PhysRevC.104.014310}.
   The quenching factors $R_s$ as a function of $\Delta S$ with and without the core screening are shown in (b).
   }
	\label{Fig3}%
\end{figure}

For the screening factor $f$ in studies of knockout reactions of carbon isotopes, the parameters are optimized to be $\alpha=0.95$ and $\gamma=3.52$, respectively.
The total densities $\rho(r)$  are calculated with the Skyrme Hartree-Fock-Bogoliubov method, using the HFBRAD solver~\cite{hfbrad}.
As shown in Fig. \ref{sf}(a),
the inner part is almost completely screened in the range of $r <$ 2 fm. The core screening effect is washed out for $r >$ 5 fm.
We also tested the core screening with another density dependent formalism $f^{'}$, which is different in the inner part but the resulting cross sections are almost unchanged (see Fig.\ref{Fig3}).
It means that the cross section is sensitive to the outer part of the wave functions in the range of 2$< r <$ 4 fm  and relatively insensitive to its inner part, showing weak dependence on the fine details of the inner part of the wave function,
as also been pointed out in Ref.~\cite{Hebborn2}.
This could be related to the low survival probability of the core when the distance $r$ is small.
 This finding is similar to the conclusions derived from studies of breakup reactions based on the Ichimura-Austern-Vincent(IVA) formalism\cite{PhysRevC.108.024606}.

To illustrate the role of core screening, the suppression of wave functions of protons and neutrons in $^{15}$C is shown in Fig. \ref{sf}(b).
To simply the analysis, the wave function of the removal nucleon is expressed as $|\psi\rangle=\sum_{nlj}C^2S(nlj)|\psi_{nlj}|^2$ ,where $\psi_{nlj}$ represents the single-particle basis states.
We see that the location of wave function of the removal neutron is more extended than that of the removal proton.
For the neutron-rich $^{15}$C, the proton orbital is deeply bound.
It can be seen that the core screening effect is more significant in the suppression of wave function of the removal proton compared to that of neutron, although both are suppressed.

%其中实线表示未考虑到屏蔽之前的约化波函数,虚线则表示考虑屏蔽因子之后的约化波函数

%单核子敲出截面的理论计算值,$\sigma_{th}^{eikonal}$表示未修正前的理论计算截面,$\sigma_{th1}^{scr},\sigma_{th2}^{scr}$分别表示使用$f_1,f_2$ 计算出来的反应截面,实验数据取自文献[],$S_i$ 表示相应敲出粒子的分离能,$\Delta S$则反应核的同位旋不对称性

To calculate the cross sections,
the spectroscopic factors  of carbon isotopes adopted in this work are based on the shell model calculations  using the WBP interaction, as described in \cite{PhysRevC.104.014310,PhysRevC.110.014603}.
The knockout reactions of carbon isotopes have been extensively studied,  providing an ideal testing ground.
The single-particle wave functions in Eq.(\ref{eq:str})  are calculated using the Woods-Saxon potential.
These parameters of Woods-Saxon potential are constrained by the root mean square (rms) radius of the wave function and the separation energy\cite{PhysRevC.77.044306,dypang}, ensuring that the rms radius of the wave function $r_{sp}$ is related to the Hartree-Fock calculated orbital radius $r_{HF}$ by the equation $r_{sp}=\sqrt{\frac{A}{A-1}}r_{HF}$ for each projectile.
The parameterization of $\sigma_{NN}$ for experimental energies is adopted from reference \cite{PhysRevC.41.1610} and the differential cross section is isotropically averaged.
To simplify the calculations, we have focused on the impact of the core screening on the stripping cross section. In order to compare with experimental data, we will proportionally reduce the contribution of diffraction dissociation cross-sections, which are much smaller compared to the stripping cross sections.

The calculated knockout cross sections of carbon isotopes are shown in Fig. \ref{Fig3}(a). For neutrons, $\Delta S$ that characterizes the isospin asymmetry of the projectile nucleus is defined as $S_n+E^*-S_p$. The average
$\Delta S$  is weighted by various single-particle cross sections.
We see the descriptions of knockout cross sections are much improved by including the core screening effect.
Different core screening formulism result in very close cross sections.
The quenching factors of the  cross section, $R_s$, are shown in Figure \ref{Fig3}(b).
We found that due to the core screening effect, the dependence of the reduction factor $R_s$ on isospin asymmetry $\Delta S$ is significantly reduced. In the absence of core screen, the obtained slope of the relationship between $R_s$ and $\Delta S$ is $-0.010\  \text{MeV}^{-1}$. We also see that $R_s$ of neutron removal from $^{14}$C is overestimated by our approach.
Actually the experimental cross sections of neutron removal from $^{14}$C and $^{16}$C are close in systematic trends although the associated $\Delta S$ are different.
In this respect, experimental uncertainties have to be taken into account~\cite{PhysRevLett.131.212503}.
 Our results demonstrated that  the core screening effect  leads to a significant reduction of isospin dependence of $R_s$ on $\Delta S$, which has been a long standing puzzle.
This highlights the crucial role of the in medium effects between the core and target nucleus.
In the future, the core screening effect is expected to be further improved, such as by considering the double-folding of the densities of target and core nuclei, to describe various knockout reactions.

%%\label{}

\section{Discussions}

In summary, we proposed the core screening effect in calculations of the knockout cross sections involving heavy-ion targets.
The core screening effect is a result of in-medium effect and Pauli blocking effect, which would suppress
the knockout cross sections of deeply bound states significantly. Consequently, the systematic
quenching of cross sections with increasing $\Delta S$ can be explained, which has been a long standing puzzle.
This finding provides new insights into the knockout reactions involving two composite systems, highlighting the finite-size effect of the reaction probe.

For the (p,pN) reactions, the core would also affect the removal of the deeply bound nucleons, but the proton has a much smaller size than  heavy ions.
Compared to knockout reactions induced by heavy ions, the core screening effect due to the proton target should be much reduced.
Therefore the quenching of spectroscopic factors as a function of isospin asymmetry has not been observed in (p,pN) reactions\cite{HOLL2019682,PASCHALIS2020135110}.
Besides, it seems that the influence of deeply bound nucleons rather than short-range correlations is more relevant for knockout reactions, since there is no strong incident energy dependence in the $R_s$
 values within a wide energy range~\cite{cpc}.

The core screening effect indicates the complex interaction between the core, the target and the removal nucleon.
The structures of the core and the target should be invoked.
It is understandable that it is more difficult for the heavy ion target to ``see" the deeply bound nucleons in the projectile due to the core screening effect.
Actually the core screening effect is not  conflicting with the core destruction effect.
It was shown recently that the beyond mean-field effect is important in heavy ion fusion reactions~\cite{kyle}.
Theoretically, it is possible but very difficult to derive the in-medium and Pauli blocking effects from first principles,
but deserves for future studies.

\acknowledgments
We thank for useful discussions with C.J. Lin and F.R. Xu.
 This work was supported by  the
 National Key R$\&$D Program of China (Grant No.2023YFA1606403, 2023YFE0101500),
  the National Natural Science Foundation of China under Grants No.12475118, 12335007.

\end{document}